# Structure of liquid and glassy methanol confined in cylindrical pores

Denis Morineau[1,2], Régis Guégan[2], Yongde Xia[1,3] and Christiane Alba-Simionesco[1]

[1]Laboratoire de Chimie Physique, CNRS-UMR 8000, Bâtiment 349,
Université de Paris-Sud, F-91405 Orsay, France

[2]Groupe Matière Condensée et Matériaux, CNRS-UMR 6626, Bâtiment 11A,
Université de Rennes 1, F-35042 Rennes, France

[3]School of Chemistry, The University of Nottingham, University Park, Nottingham
NG7 2RD, United Kingdom

## ABSTRACT

We present a neutron scattering analysis of the density and the static structure factor of confined methanol at various temperatures. Confinement is performed in the cylindrical pores of MCM-41 silicates with pore diameters $D$=24 Å and $D$=35 Å. A change of the thermal expansivity of confined methanol at low temperature is the signature of a glass transition, which occurs at higher temperature for the smallest pore. This is an evidence of a surface induced slowing down of the dynamics of the fluid. The structure factor presents a systematic evolution with the pore diameter, which has been analyzed in terms of excluded volume effects and fluid-matrix cross-correlation. Conversely to the case of Van der Waals fluids, it shows that stronger fluid-matrix correlations must be invoked most probably in relation with the H-bonding character of both methanol and silicate surface.

## I. INTRODUCTION

During the last decade, an abundant literature reports on the physical properties of fluids confined in various environments. Some leading issues include the phase transition behavior,[1] glass transition and dynamical properties of molecular liquids in nanometer-scale porous geometry.[2] In this respect, it is also required to elucidate how structural features are modified by confinement effect and at the fluid-substrate interface.

Some of these aspects have already been addressed by different groups. It has been recently shown that confinement of a weakly interacting fluid in porous silicate may significantly affects its static properties, leading to glassy phases with different density at low temperature.[3] Moreover, the intermolecular correlations related to a preferential local structure within the fluid are probably affected by confinement[4] and peculiar features such as contact layers or orientation ordering have been invoked for some confinement conditions.[5,6,7] Confinement could eventually lead to new phases, the formation of which is driven by the balance between fluid-fluid and fluid-substrate interaction.[8] In the case of poor fluid-wall interaction, we have demonstrated that the change with pore size of the experimental structure factor of a confined fluid is dominated by topological constraints in terms of excluded volume effects.[9] The strongest fluid-substrate interaction has been mostly addressed in the case of water confined in various porous silicates by diffraction experiments and molecular simulation, [4,6,10,11] where strong distortions of the hydrogen-bond network of liquid water are reported. It is established that some hydrogen-bond bridges with the hydrophilic surface play a major role in defining the structure of the confined fluid.

Alcohols are strongly hydrogen-bonded organic liquids. Methanol being the simplest one is of fundamental interest and a test system for studying the influence of H-bonds and fluid-substrate interaction on the properties of confined fluids. Its structural properties have been extensively studied in bulk conditions by neutron scattering,[12,13] X-rays[14] and molecular simulation.[15,16] Phase transitions of methanol confined in regular cylindrical pores of MCM-41 and SBA-15 silicates have been investigated by X-rays diffraction.[17] The freezing/melting behavior depends markedly upon the pore size, with different degrees of hysteresis and metastability. Within the pores of diameter lower than 78 Å, crystallization never occurs and the confined methanol ultimately vitrifies at about 100 K close to the value of the bulk glass transition temperature.

The aim of our study is to investigate the local structure within the amorphous phases of methanol (liquid and glass) confined in a cylindrical pore of MCM-41 silicate. MCM-41 is regarded as one of the most regular mesoporous materials currently available. It presents cylindrical-like highly monodisperse parallel pores arranged in a triangular crystalline lattice. Neutron scattering experiments have been performed in order to probe how the average density and the intermolecular structure factor of methanol change with temperature and the pore size.

## II. EXPERIMENTAL SECTION

### A. Materials

Two kinds of MCM-41 with pore diameters of 24 and 35 Å were synthesized according to a procedure already detailed in previous works.[3,9,18] Decyl- and hexadecyl- ammonium bromide were respectively used as template in order to vary the pore diameter. The structural parameters of the matrices were checked by neutron scattering (cf. Fig. 1) and by nitrogen adsorption isotherms at liquid nitrogen temperature.

Template organics were removed by calcination in the air at 823 K and the remaining surface silanol groups were deuterated by isotopic substitution with heavy water. The same procedure as described in refs. 3,9 was used to achieve a nearly complete filling of the porous volume with methanol. The loaded cells were sealed with an indium joint with no detectable indication of loss of methanol or contamination with water during the experiments. Fully deuterated methanol (D=99.8%) was purchased from Eurisotop, Saclay.

## B. Measurements

The neutron scattering experiments were performed using the two spectrometers G6.1 and 7C2 at the Orphée reactor neutron source of the Laboratoire Léon Brillouin.[1] These two double-axis spectrometers are based on a similar device. Their main difference is the use of a cold and a hot neutron source, respectively. This enables to cover an extended $Q$-range from 0.1 Å$^{-1}$ to 10 Å$^{-1}$ by selecting different monochromatic incident wavelengths (4.7 Å and 1.1 Å).

A thin walled cylindrical vanadium cell with an inner diameter of 8.0 mm and a height of 80 mm was used. It was mounted on a liquid Helium cryostat. A standard correction procedure has been applied to the experimental intensities to take into account the detector efficiencies, backgrounds, attenuation, multiple scattering. Normalization to absolute units and Placzek corrections have been applied to the spectra obtained on 7C2 from a fit of the intramolecular contribution to the high-$Q$ part (see refs. 9 and 19 for details).

## III. DENSITY MEASUREMENTS

The direct measurement of the density of a fluid confined in MCM-41 materials has previously been achieved for confined benzene and toluene.[3,20] This method uses the crystalline nature of the underlying arrangement of the cylindrical pores. Figure 1 presents the differential cross section of an empty matrix with two different experimental resolutions for transfer of momentum $Q$ smaller or larger than 1 Å$^{-1}$. Above 1 Å$^{-1}$, one recovers the structure factor of the amorphous silica that forms the walls. At smaller $Q$ values, the crystalline superstructure of the porous geometry is reflected by sharp Bragg peaks. The intensity of these peaks is related to the square of the contrast, the contrast being defined as the difference of scattering length density between the silica matrix and the pore content. It is written as

$$I = A\left( \rho_{SiO_2} \overline{b_{SiO_2}} - \rho_{MeOD} \overline{b_{MeOD}} \right)^2 \qquad (1)$$

where A is a constant term, $\overline{b}$ is the average coherent scattering length and $\rho$ is the number density of $SiO_2$ or methanol (MeOD); in the empty pore, $\rho_{MeOD} = 0$. Increasing the average scattering length density of the confined fluid, by a gradual adsorption, by using an H/D mixture or by changing the temperature, the intensity of the Bragg peaks behaves as schematically shown in Fig. 2. It firstly decreases, vanishes when the exact contrast matching condition is fulfilled and increases again. Contrast matching experiments have been performed previously with an H/D mixture to validate this method.[3] In the present work, we have used matrices completely loaded with fully deuterated methanol. The scattering length density of condensed methanol is always larger than that of silica so that the intensity of the Bragg peak monotonously increases as the density of the confined fluid.

---

[1] Laboratoire Léon Brillouin CEA-CNRS, Saclay, France.

The differential cross section of methanol confined in the two MCM-41 for a temperature change ranging from 20 K to 300 K is shown in Figs. 3(a) and 3(b). The 100, 110 and 200 Bragg peaks are observed but the temperature dependence is more easily monitored from the first and most intense one. For both samples, a huge increase of the intensity of this peak is observed as the temperature decreases. This method is in fact extremely sensitive to any subtle change of density within the confined phase. From 300 to 10 K, the change in density of methanol is about 18%. Meanwhile, the intensity of the Bragg peak increases by a factor of 2.7. Moreover, the scattered intensity at the Bragg peak location is very large so that an excellent statistics is rapidly obtained. As an indication, the maximum intensity of the 100 Bragg peak is more than 10 times larger than the intensity at the maximum of the structure factor of methanol at $Q \sim 1.8$ Å$^{-1}$.

Lowering the temperature, the intensity of the 100 Bragg peak increases then bends around 100-130 K. This feature is observed for the two pore diameters. It indicates that the thermal expansivity of the confined methanol changes rather abruptly in this temperature range. This is recognized as the signature of a glass transition of the confined phase.

From Eq. (1), one can extract the density of the confined phase from a combination of the intensities of the 100 Bragg peaks of the filled and the empty matrices. The density of the silica wall firstly used in the calculation was the same as the one of amorphous silica at ambient pressure (2.2 g.cm$^{-3}$). A value of the same order is expected from the wide-angle structure factor of the empty matrices. From this direct calculation, the density of confined methanol looks systematically smaller than the bulk one by about 8%. Such a difference is however not consistent with the wide angle neutron scattering experiments presented below and should probably be disregarded. A possible error of a few percents due to either the silica wall density, the amount of MCM-41 under the neutron beam or the loading of methanol can not be excluded. This induces a systematic error in the absolute value of the methanol density but does not affect at all its relative temperature dependence. Therefore, the density of the confined phase has been rescaled to the bulk one at high temperature. Taking the density of the silica wall equal to 2.02 g.cm$^{-3}$ would have the same effect as this scaling procedure. The obtained results are shown in Fig. 4. Two temperature regions are observed. Above 125 K, the density of the confined methanol depends almost linearly on the temperature. Its temperature derivative for the two pore diameters is in good agreement with the thermal expansivity of the bulk liquid (1.0 10$^{-4}$ to 1.1 10$^{-4}$ K$^{-1}$). At lower temperature, the slope is much smaller with thermal expansivity coefficients values (8.3 10$^{-5}$ K$^{-1}$ and 10 10$^{-5}$ K$^{-1}$ for $D$=24 Å and $D$=35 Å, respectively) which are typical for a solid phase. This is the signature of a glass transition. Crystallisation can be ruled out since no Bragg peaks from methanol crystallites are observed at wide angles, in agreement with previous results.[17]

Crystallisation of bulk methanol cannot be avoided at low temperature so that the glassy state is unreachable with usual quenching rates. However, a glass transition has been reported for methanol deposited on a cold substrate from its vapour phase.[21] The calorimetric glass transition has been located at about 103K (denoted as bulk Tg in Fig. 4). From the bend of the density, it is possible to estimate the temperature region where the glass transition of the confined methanol occurs. It is 110 K ±10 K for a pore diameter $D$=35 Å, thus very close to the bulk one. For a pore diameter $D$=24 Å, the glass transition at 120 K ±10 K is unambiguously larger than the bulk one. An increase of the glass transition temperature of liquids confined in small pores has already been observed for toluene confined in MCM-41[3] and polypropylene glycol confined in CPG's.[22] It has been attributed to surface dominant effects, which induce a strong slowing down of the structural relaxation at the fluid-wall surface boundary.

Morishige et al. have located the glass transition of methanol confined in various MCM-41 from the $Q$-location of the maximum $Q_{max}$ of its main diffraction peak.[17] They have reported that the glass transition was around 100 K for the four pore diameters ranging from 78 Å to 24 Å. Their result do not exclude a possible increase of the glass transition of confined methanol for smallest pores. Indeed, only one of the four MCM-41 used in that work ($D$=24 Å) has a size which compares with the smallest pore size we used and the results may depend on the definition of the pore diameter. BJH method, BET volume to surface ratio or diffraction techniques yields slightly different values of $D$. In our study, a recommended method combining diffraction and adsorption has been applied.[23] But above all, the method we developed to detect Tg presents an improved sensitivity. The main diffraction peak position $Q_{max}$ of a liquid is roughly associated to the density of the liquid and only barely changes with it :in the whole temperature range, a change of less than 10% of $Q_{max}$ is observed whereas here the intensity of the 100 Bragg peak changes by 170%.

## IV. STRUCTURE FACTOR

Diffraction is a unique experimental tool to investigate the local structure of a fluid. The formalism used to relate the structure factor of molecular liquids to the differential cross section measured by neutron scattering has been extensively documented. Only essential concepts will be recalled here and more details are offered to the reader in specialized publications.[24]

For bulk methanol, the total structure factor of $N_{MeOD}$ liquid molecules is given by

$$S_M(Q) = \frac{1}{N_{MeOD}\hat{b}_{MeOD}^2} \left\langle \sum_i^{MeOD} \sum_j^{MeOD} b_i b_j \exp(i\mathbf{Q} \cdot \mathbf{r}_{ij}) \right\rangle_{\theta,\varphi} \tag{2}$$

where $\hat{b}_{MeOD}$ is the sum of the coherent scattering lengths relative to the atoms constituting one molecule. $\langle\ \rangle_{\theta,\varphi}$ stands for an average over the angles of $\mathbf{Q}$ in spherical coordinates of a sum of the scattering intensity arising from every couples of atoms in the sample. As shown in Fig. 5, it is convenient to split $S_M(Q)$ into an intramolecular form factor $f_1(Q)$ and an intermolecular contribution $D_M(Q)$. It allows calculating the intermolecular pair correlation function $g_L(r)$ by Fourier transform according to

$$S_M(Q) = f_1(Q) + \frac{4\pi}{Q}\rho_M \int (g_L(r)-1)r\sin(Qr)dr \tag{3}$$

where $\rho_M$ is the molecular density of the liquid. As shown in Fig. 5, the structure factor at large $Q$, above 3Å$^{-1}$ is essentially related to intramolecular contributions. Most of the information about the intermolecular short range order is included in the $Q$-range of the so-called main diffraction peak (below 4 Å$^{-1}$). The radial distribution function (cf. inset Fig. 5 in the form r$^2$(g(r)-1)) illustrates the extension of intermolecular correlations up to several molecular diameters, which is promoted by H-bonds.

In order to study how the structure of methanol is altered in confined geometry, one must address several difficulties. The experimental differential cross section is now the sum of three terms according to Eq. (4).

$$F^{\text{filled}}(Q) = X_{\text{MeOD}} \hat{b}_{\text{MeOD}}^{\ 2} S_M^{\ \text{MeOD}}(Q) + X_{\text{SiO}_2} \hat{b}_{\text{SiO}_2}^{\ 2} S_M^{\ \text{SiO}_2}(Q) + 2\sqrt{X_{\text{MeOD}} X_{\text{SiO}_2}} \hat{b}_{\text{MeOD}} \hat{b}_{\text{SiO}_2} S_M^{\ \text{SiO}_2-\text{MeOD}}(Q)$$

$$(4)$$

where $X_i=N_i/N$ is the fraction of molecules of type $i$ in the sample composed by $N_{\text{MeOD}}$ methanol molecules confined in a matrix of $N_{\text{SiO}_2}$ silica units. The three partial structure factors are defined as [25]

$$S_M^{\ \text{MeOD}}(Q) = \frac{1}{N_{\text{MeOD}} \hat{b}_{\text{MeOD}}^{\ 2}} \sum_i^{\text{MeOD}} \sum_j^{\text{MeOD}} b_i b_j \left\langle \exp(i\mathbf{Q} \cdot \mathbf{r}_{ij}) \right\rangle_{\theta,\varphi} \qquad (5)$$

$$S_M^{\ \text{SiO}_2}(Q) = \frac{1}{N_{\text{SiO}_2} \hat{b}_{\text{SiO}_2}^{\ 2}} \sum_i^{\text{SiO}_2} \sum_j^{\text{SiO}_2} b_i b_j \left\langle \exp(i\mathbf{Q} \cdot \mathbf{r}_{ij}) \right\rangle_{\theta,\varphi} \qquad (6)$$

$$S_M^{\ \text{MeOD}-\text{SiO}_2}(Q) = \frac{1}{\sqrt{N_{\text{MeOD}} N_{\text{SiO}_2}} \hat{b}_{\text{Bz}} \hat{b}_{\text{SiO}_2}} \sum_i^{\text{MeOD}} \sum_j^{\text{SiO}_2} b_i b_j \left\langle \exp(i\mathbf{Q} \cdot \mathbf{r}_{ij}) \right\rangle_{\theta,\varphi} \qquad (7)$$

The two first terms are related to the interference of neutrons scattered by the matrix and the confined fluid by themselves whereas the third one is related to the cross correlation between the matrix and the methanol. The matrix-matrix correlations are directly evaluated by using the experimental differential cross section of the empty matrix. As shown in Figs. 6(a) and 6(b), this contributes to about one quarter to one third of the total differential cross section. The composite scattering function $S(Q)$ is finally obtained after subtraction of the matrix-matrix correlations and normalization to one methanol molecule (see ref. 9 for details). $S(Q)$ is thus the sum of methanol-methanol and methanol-silica correlations which can not be easily separated. Moreover, both contributions are affected by the so-called exclusion volume effect. This effect relates the fact that pair correlation functions of a confined phase are sensitive to both the local structure of the fluid ('intrinsic intermolecular correlations') and the requirement that a fraction of space is inaccessible to the molecules (out of pore volume). It has been recently addressed in the literature for water confined in a vycor glass[10] and benzene in MCM-41 and SBA-15 mesoporous silicates.[9] In this latest case, it demonstrates that the strong distortions of the experimental structure factors of confined benzene are essentially due to an excluded volume effect. Differences in the local structure of the confined fluid (intrinsic fluid-fluid correlations) and the occurrence of non-homogeneous benzene-matrix correlations are therefore hardly quantified experimentally. This is probably enhanced with benzene for which the fluid-fluid correlations in bulk and fluid-wall interactions are weak. In the case of methanol, the excluded volume effect has been evaluated from the bulk pair correlation function and from the uniform pair correlation functions that only depend on the topology of the porous volume, which is precisely characterized for MCM-41. The structure factor can be written as follows if the confined phase is homogeneous, which is the case for a liquid. [9]

$$S(Q) = f_1(Q) + X_{\text{MeOD}} \frac{4\pi}{Q} \rho \int ((\tilde{g}^{\text{MeOD}}(r) - 1) g_u^{(pp)\text{intra}}(r) + g_u^{(pp)}(r) - 1) r \sin(Qr) dr$$

$$\left[ + 2\frac{\hat{b}_{SiO_2}}{\hat{b}_{MeOD}} X_{SiO_2} \frac{4\pi}{Q} \rho \int (\tilde{g}^{MeOD-SiO_2}(r) g_u^{(pw)}(r) - 1) r \sin(Qr) dr \right. \tag{8}$$

$f_1(Q)$ reflects intramolecular contributions. $\tilde{g}^{MeOD}(r)$ and $\tilde{g}^{MeOD-SiO_2}(r)$ reveal the intrinsic correlations within the fluid and between the fluid and the matrix. $g_u^{(pp)intra}(r)$, $g_u^{(pp)}(r)$ and $g_u^{(pw)}(r)$ are different uniform fluid pair correlation functions, also defined as the pair correlation function of a system of noninteracting particles confined in the same restricted geometry. They reflect respectively the intrapore pore-pore correlations, the pore-pore correlations and the pore-wall correlations. They only depend on the geometry of the porous matrix and a method of computation for MCM-41 types of porous geometry is fully detailed in ref. 9.

The temperature evolution of the experimental structure factor $S(Q)$ is shown in Fig. 7. The bulk methanol does not supercool but crystallizes on cooling for usual quench rates so that the equivalent spectrum at 70 K is not presented for the bulk. On the contrary, supercooling and vitrification is possible in confined geometry. The temperature evolution of the structure factor of methanol is qualitatively the same for the bulk and the two pore diameters. It essentially affects the main diffraction peak (around $Q=1.8$ Å$^{-1}$), which shifts to larger momentum and narrows as the temperature decreases. This has commonly been attributed to the increase of local packing and the improvement of the ordering in terms of short nonlinear chains induced by H-bonds.[12]

In order to achieve a description of the effect of confinement on the structural organization of methanol, one needs to evaluate quantitatively the additional contributions arising from methanol-matrix cross correlations and excluded volume effects. Doing so, we have been able to reproduce the experimental results with a model based on simple assumptions. In the following analysis, the extent of excluded volume effects and cross correlation terms are computed assuming that the confined liquid presents the intrinsic short range order of the bulk. Therefore, the experimental methanol-methanol composite pair correlation function $g_L(r)$ of bulk liquid at 290 K stands in for $\tilde{g}^{MeOD}(r)$ in Eq. (8). In the case of weak interactions between the fluid and the matrix the liquid-silica cross correlation function $\tilde{g}^{MeOD-SiO_2}(r)$ could be approximated to unity. It has been proved to be a good approximation for benzene but not for methanol. This is certainly due to H-bonds interactions between methanol and the silica surface and the silanol groups. It is necessary to represent the methanol-silica cross correlations with a more realistic function. This last function $\tilde{g}^{MeOD-SiO_2}(r)$ has been extracted from a molecular simulation study of confined methanol, which will be reported in a nearest article.[26] It is sufficient to mention here that this work is based on a canonical Monte Carlo simulation of 229 molecules confined in a cylindrical pore of diameter 24Å created within a modeled silica glass. We have followed the procedure proposed by Brodka et al.[5] to carve a cylindrical cavity within an equilibrium cubic structure of amorphous silica of 36Å on a side provided by Vink et al.[27] In this case, nonbridging oxygens are saturated with hydrogen atoms to form surface hydroxyl groups. Although the silica matrix is subsequently kept rigid, rotation around the Si-O bond of the hydroxyl groups is allowed. OPLS interaction potential has been used for methanol-methanol interactions.[15] Interaction potential parameters from ref. 5 for silica and Lorentz-Berthelod mixing rules have been applied to model the methanol-silica interactions. This procedure leads to a realistic description of the irregular inner surface of the porous silicate and of the interfacial interactions between the fluid and the matrix.

The excluded volume effect on these two pair correlation functions (experimental $g_L(r)$ for methanol-methanol and simulated $\tilde{g}^{MeOD-SiO_2}(r)$ for methanol-silica) has been computed using the pore-pore $g_u^{(pp)}(r)$ and pore-wall $g_u^{(pw)}(r)$ uniform pair correlation functions computed for an array of cylindrical pores according to ref. 9, which models the experimental porous geometries as shown in Fig. 8.  Under these assumptions, Eq. (8) is written as

$$S(Q) = f_1(Q) + X_{\text{MeOD}} \frac{4\pi}{Q} \rho \int (g_L^{\text{MeOD}}(r) - 1) g_u^{(pp)\text{intra}}(r) r \sin(Qr) dr + X_{\text{MeOD}} \frac{\rho}{\rho_p} S_u^{(pp)}(Q)$$

$$\underbrace{\phantom{f_1(Q)}}_{\text{intramolecular}} \qquad \underbrace{\phantom{X_{\text{MeOD}} \frac{4\pi}{Q} \rho \int (g_L^{\text{MeOD}}(r) - 1) g_u^{(pp)\text{intra}}(r) r \sin(Qr) dr}}_{\text{intrapore}} \qquad \underbrace{\phantom{X_{\text{MeOD}} \frac{\rho}{\rho_p} S_u^{(pp)}(Q)}}_{\text{interpore}}$$

$$- 2 \frac{\hat{b}_{\text{SiO}_2}}{\hat{b}_{\text{MeOD}}} X_{\text{SiO}_2} \frac{\rho}{1 - \rho_p} S_u^{(pp)}(Q) + 2 \frac{\hat{b}_{\text{SiO}_2}}{\hat{b}_{\text{MeOD}}} X_{\text{SiO}_2} \frac{4\pi}{Q} \rho \int (\tilde{g}^{\text{MeOD-SiO}_2}(r) - 1) g_u^{(pw)}(r) r \sin(Qr) dr$$

$$\underbrace{\phantom{- 2 \frac{\hat{b}_{\text{SiO}_2}}{\hat{b}_{\text{MeOD}}} X_{\text{SiO}_2} \frac{\rho}{1 - \rho_p} S_u^{(pp)}(Q)}}_{\text{uniform cross correlation}} \qquad \underbrace{\phantom{2 \frac{\hat{b}_{\text{SiO}_2}}{\hat{b}_{\text{MeOD}}} X_{\text{SiO}_2} \frac{4\pi}{Q} \rho \int (\tilde{g}^{\text{MeOD-SiO}_2}(r) - 1) g_u^{(pw)}(r) r \sin(Qr) dr}}_{\text{non uniform cross correlation}}$$

$$(9)$$

Eq. (9) is an extension of the previously published analysis of confined benzene, (see Eq. (19) of ref. 9), which now expresses the liquid-silica cross correlations into two terms : the uniform correlation is the approximation of 1$^{\text{st}}$ order obtained taking $\tilde{g}^{MeOD-\text{SiO}_2}(r)$ equals to one and directly computed in the reciprocal space. The non-uniform correlation is the correction to this approximation, which introduces non-trivial liquid-silica correlation ($\tilde{g}^{MeOD-\text{SiO}_2}(r) - 1$) extracted from molecular simulation results. The various contributions are shown in Fig. 9 for a pore diameter $D$=24 Å.

Three essential features are observed by the computation, which reproduces the experimental observations (also shown in Figs. 10(a) and 10(b)). First is a large negative contribution in the low-$Q$ limit, attributed  to the uniform cross correlation term. This feature reflects the reduction of the intensity of the Bragg peaks of the MCM-41 by contrast effects when filled with a liquid as discussed in the previous section. The second feature is a decrease of intensity and broadening of the main diffraction peak of methanol (from about 1.2 to 2.2 Å$^{-1}$ ). It is essentially attributed to the intrapore excluded volume effect. It reflects how the topological constraint of two molecules being within the same cylindrical pore affects their pair correlations. The last features observed experimentally are a shift of the main diffraction peak to larger $Q$-values and a slight decrease of the intensity of the second maximum (around $Q=3$ Å$^{-1}$) (See arrows and solid guideline in Fig. 10(a)). These last experimental features cannot be reproduced computationally if only uniform liquid-silica correlations are assumed. Indeed recent molecular simulations prove that significant methanol-silica interactions occur at the liquid-solid interface (see F ig. 8). The non uniform cross correlations in $Q$-space, as shown in dashed line in Fig. 9, present a double oscillation within the $Q$-range of the main diffraction peak of methanol. This particular shape induces an apparent shift of the position of the main diffraction peak, which is confirmed experimentally.

Important changes in the structure factor of another H-bonding glass-forming system, the *m*-toluidine, confined in a MCM-41 have also been observed experimentally (cf. Fig. 11).[29] It particularly concerns the shape of the main diffraction peak around 1.3Å$^{-1}$. This is obviously an additional example where fluid-wall H-bond interactions play a major role and should be confirmed by a more quantitative analysis of the fluid-matrix correlations.

## V. CONCLUSIONS.

Fluids in nanometer scale geometry are the subject of active researches in various fields, including catalysis, nanotribology,  biology or fundamental physics. The understanding of the static properties of the confined phase, *i.e.* the density, the local arrangement of the molecules within the pore and at the solid interface, currently requires much effort and remains a challenge. It is essential to address the relationship between the structure of confined molecules, their dynamical or thermodynamical properties and the kinetics of the phase transitions within the pore.

Methanol has been chosen as a test system for studying the influence of H-bonds and fluid-substrate interaction on the properties of confined fluids. Combining small and large angles neutron scattering spectrometers, it is possible to measure two essential static properties, which are the density and the static structure factor. Both measurements require a highly regular porous geometry in order to measure Bragg peaks from the porous superstructure and to compute accurately the exclusion volume effects. MCM-41 and analogous molecular sieves are currently the most convenient materials for these purposes.

Density measurements have confirmed the occurrence of a glass transition of the confined methanol, which is close to the bulk one for a pore diameter $D$=35 Å but about 20 K above for a pore diameter $D$=24 Å. The observed change in the T-dependence of the density in small pores at a glass transition temperature significantly larger than the bulk one might be quite general since it has already been observed for a non polar fluid (benzene), a polar weakly interacting Van der Waals fluid (toluene) and a H-bonding fluid (methanol).[3,20] It is also consistent with the general agreement that liquid-solid interface plays the dominant role in determining the dynamical properties of the fluid confined in small pores.[22,30,31,32]

A careful analysis of the structure factor of the confined methanol requires that excluded volume and cross correlation terms are entirely quantified. In the case of the weak interacting system benzene, excluded volume effect is the prevailing origin of the distortion of the structure factor. The additional features observed for confined methanol, such as a shift of the main diffraction peak towards higher Q, are experimental evidences of specific fluid-wall correlations most probably related to stronger H-bond interactions. These features are in agreement with the fluid-wall correlation functions recently calculated by molecular simulations. A microscopic description of the local arrangement of the molecules at the interface in terms of orientation ordering and alcohol-substrate H-bonds is required and presently under study.[28]

**ACKNOWLEDGEMENTS**

The authors are very grateful to Drs. M.-C. Bellissent-Funel and I. Mirebeau for their help in the neutron scattering experiments.

**FIGURE CAPTIONS**

**FIG. 1.** Experimental differential cross section $I(Q)$ (and $F(Q)$ in normalized units) of the MCM-41 with pore diameter of 35 Å. The left and right parts of the figure ($Q$ lower and greater than 1 Å$^{-1}$) have been measured with two different experimental resolutions and diffractometers.

**FIG. 2.** Schematic evolution of the differential cross section $I(Q)$ of an MCM-41 during the increase of the scattering length density of the confined materials. These features occur during a gradual adsorption, when the density of the fluid changes with temperature, or using H/D isotopic substitution.

**FIG. 3.** Temperature evolution of the experimental differential cross section of an MCM-41 filled with methanol with different pore diameter (a) $D$=24 Å, (b) $D$=35 Å. Here the high $Q$-resolution at small momentum transfer $Q$ allows studying the Bragg peaks of the mesoporous superstructure. The increase of the intensity of the 100 peak reflects the temperature variation of the density of the confined methanol.

**FIG. 4.** Temperature evolution of the density of the confined methanol for $D$=24 Å (open circles) and $D$=35 Å (filled circles). The bulk density is shown above the temperature of crystallization (solid line). The bulk glass transition corresponds to a vapor deposited sample (dashed line).

**FIG. 5.** Experimental structure factor (black line) and intramolecular form factor (grey line) of bulk liquid methanol at 290 K. Inset : Experimental radial distribution obtained by the Fourier transformation of the intramolecular term.

**FIG. 6.** Experimental differential cross section $F(Q)$ of the MCM-41 with different pore diameters (a) $D=24$ Å, (b) $D=35$ Å. Matrix filled with liquid methanol (black line) and the empty matrix (grey line). The mole fraction $X_{SiO_2}$ is used to directly estimate the contribution of the empty matrix to the total cross section.

**FIG. 7.** Experimental structure factor of bulk and confined methanol at different temperatures 290 K (black), 200 K (dark grey) and 70 K (grey). (a) bulk liquid. Below 200 K, the bulk methanol crystallizes, (b) D=35 Å, (c) D=24 Å.

**FIG. 8.** Methanol-methanol composite pair correlation function $g_L(r)$ of bulk methanol at 290 K obtained by neutron diffraction (thin grey line). Methanol-silica composite cross correlation function $\tilde{g}^{MeOD-SiO_2}(r)$ obtained by molecular simulation of confined methanol at 290 K (thin black line). The excluded volume effect on these Methanol-methanol and Methanol-silica pair correlation functions has been computed for a pore diameter $D=24$ Å (thick solid lines), using the pore-pore $g_u^{(pp)}(r)$ and pore-wall $g_u^{(pw)}(r)$ uniform pair correlation functions (dashed lines). (see text for details).

**FIG. 9.** Structure factor of bulk methanol at 290 K (thin solid line) and different contributions to the structure factor of confined methanol computed from the structure factor of bulk methanol at 290 K and the uniform pair correlation functions of a triangular array of cylindrical pores of diameter $D$=24 Å and wall thickness $W$=9 Å. Composite structure factor $S(Q)$ including methanol-methanol and methanol-matrix cross terms (black line). Intrapore methanol-methanol term (dark grey line). Interpore methanol-methanol term (grey line). Methanol-matrix cross term assuming no intrinsic correlations between methanol and silica (dotted line) and correction to this function if intrinsic correlations between methanol and silica obtained by molecular simulation are considered (dashed line).

**FIG. 10.** (a) Experimental structure factor of bulk methanol at 293 K (black line) and experimental composite structure factors $S(Q)$ of methanol confined in different matrices $D$=35 Å (dark grey line), $D$=24 Å (grey line). (b) Experimental structure factor of bulk methanol at 293 K (black line) and composite structure factors $S(Q)$ of methanol computed from the structure factor of bulk methanol, methanol-matrix cross correlation function from molecular simulation and the uniform pair correlation functions of a triangular array of cylindrical pores $D$=35 Å (dark grey line), $D$=24 Å (grey line).

**FIG. 11.** Temperature dependence of the experimental structure factor of another H-bonding system : the $m$-toluidine. The 3 curves correspond to temperatures of 250 K, 150 K and 110 K from top to bottom. Bulk Tg is 183.5 K (a) Bulk $m$-toluidine. The thin curve is obtained with a larger $Q$-values diffractometer (b) $m$-toluidine confined in a MCM-41 of diameter $D$=28 Å.

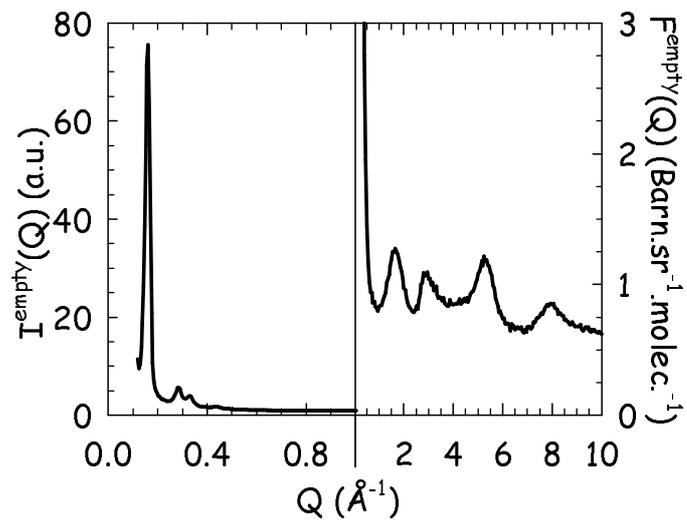

Figure 1

**Structure of liquid and glassy methanol confined in cylindrical pores.**

D. Morineau et al.

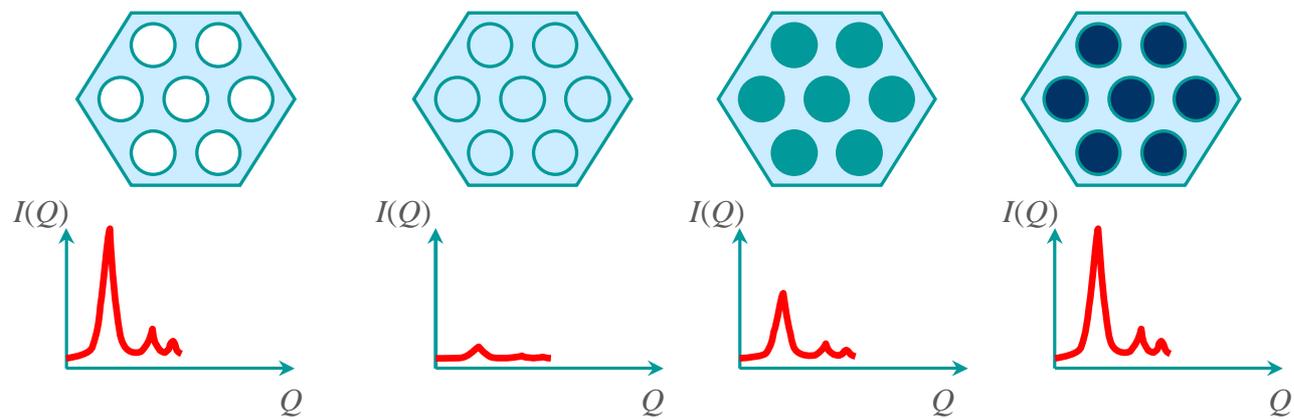

Figure 2
**Structure of liquid and glassy methanol confined in cylindrical pores.**
D. Morineau et al.

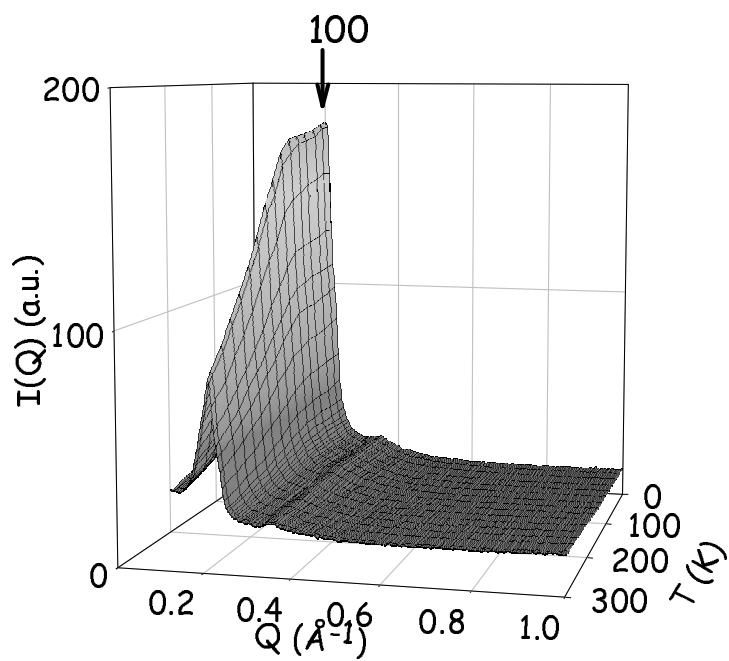

Figure 3 (a)

**Structure of liquid and glassy methanol confined in cylindrical pores.**

D. Morineau et al.

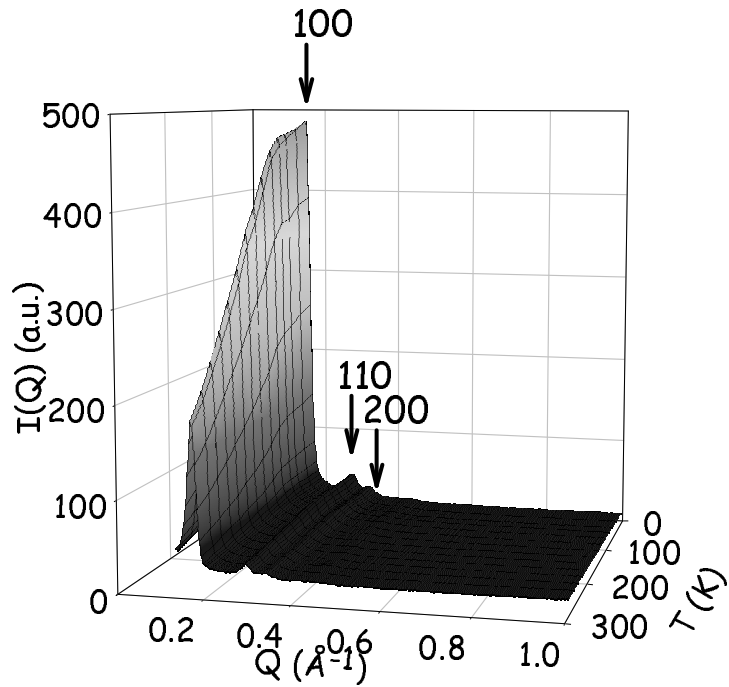

Figure 3 (b)
**Structure of liquid and glassy methanol confined in cylindrical pores.**
D. Morineau et al.

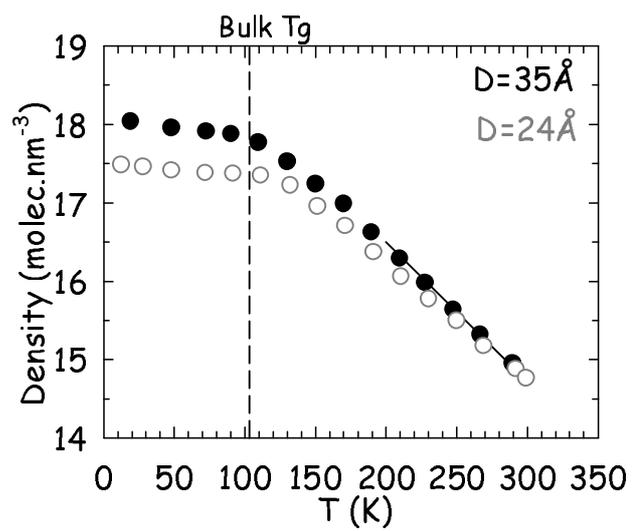

Figure 4

**Structure of liquid and glassy methanol confined in cylindrical pores.**

D. Morineau et al.

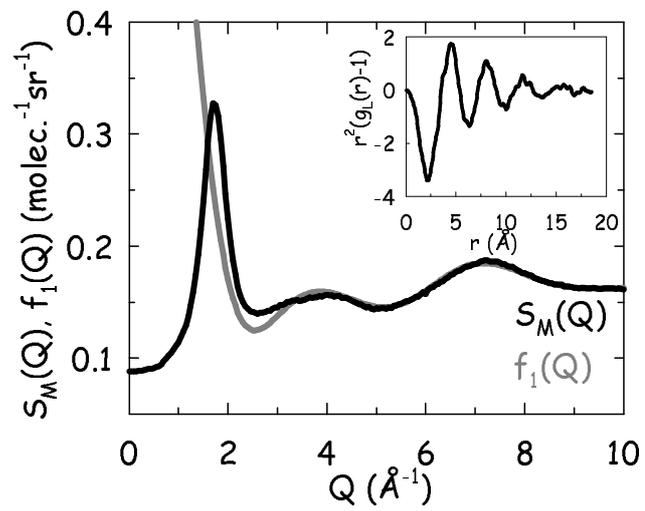

Figure 5
**Structure of liquid and glassy methanol confined in cylindrical pores.**
D. Morineau et al.

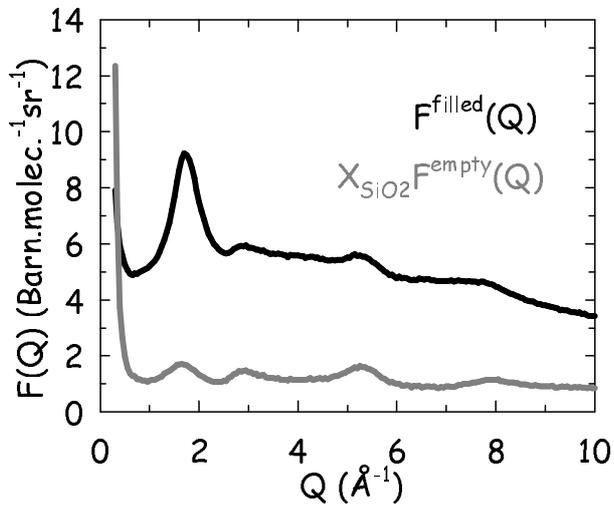

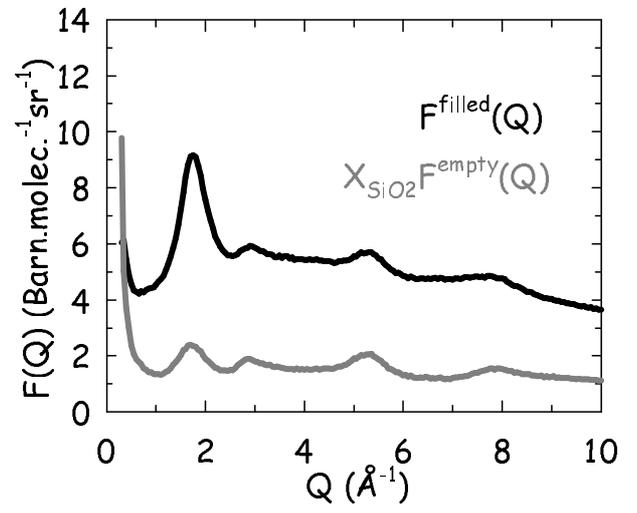

(a)                                              (b)

Figure 6
**Structure of liquid and glassy methanol confined in cylindrical pores.**
D. Morineau et al.

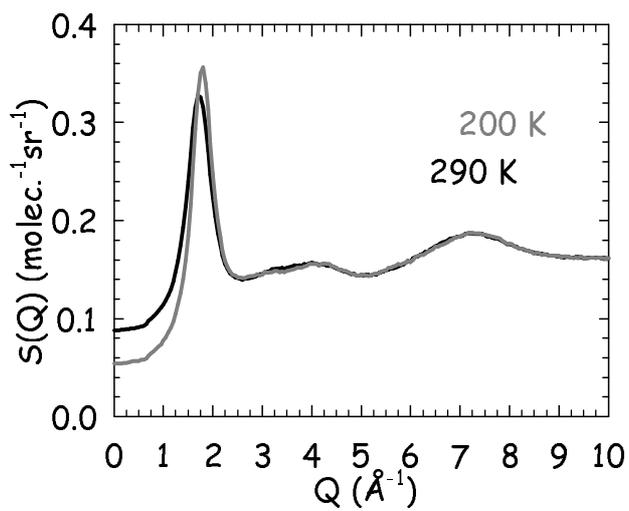

(a)

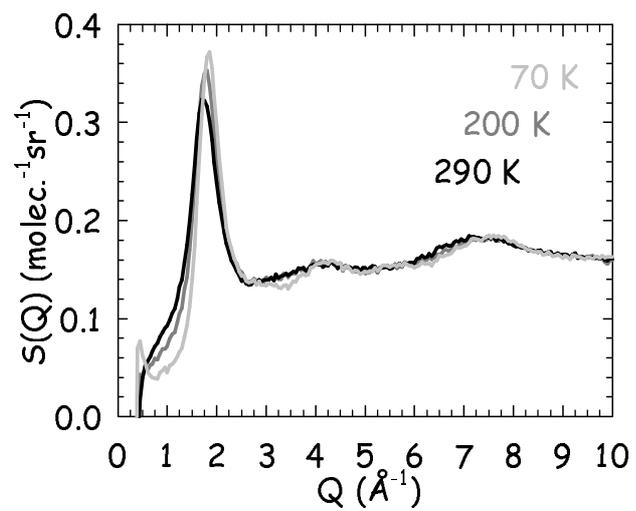

(b)

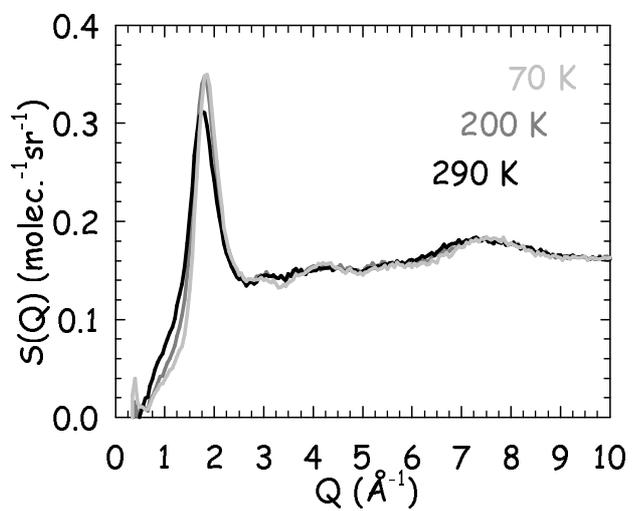

(c)

Figure 7
**Structure of liquid and glassy methanol confined in cylindrical pores.**
D. Morineau et al.

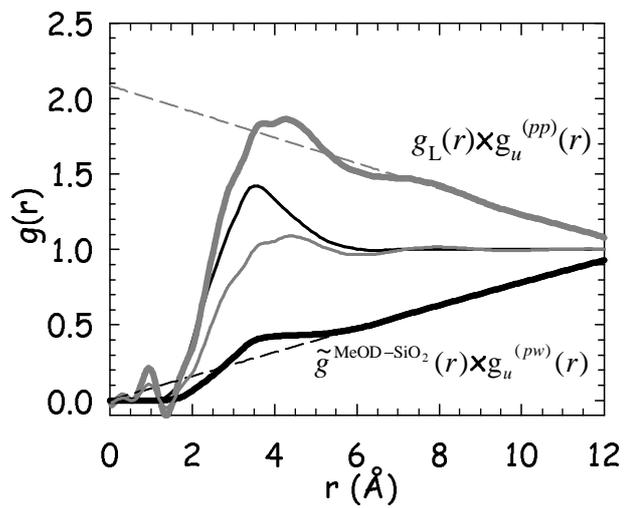

Figure 8
**Structure of liquid and glassy methanol confined in cylindrical pores.**
D. Morineau et al.

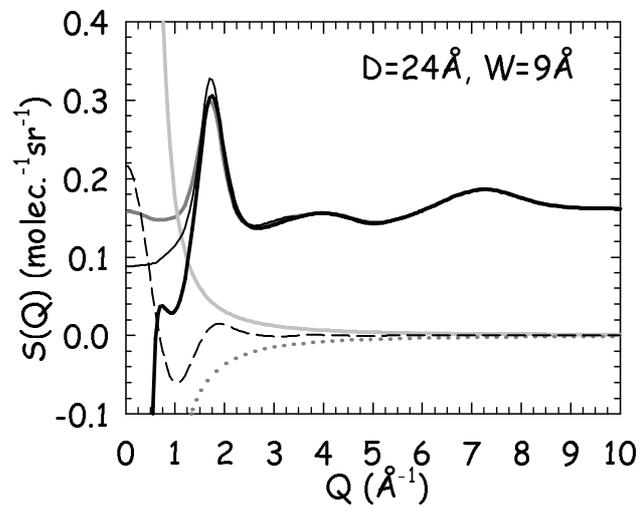

Figure 9
**Structure of liquid and glassy methanol confined in cylindrical pores.**
D. Morineau et al.

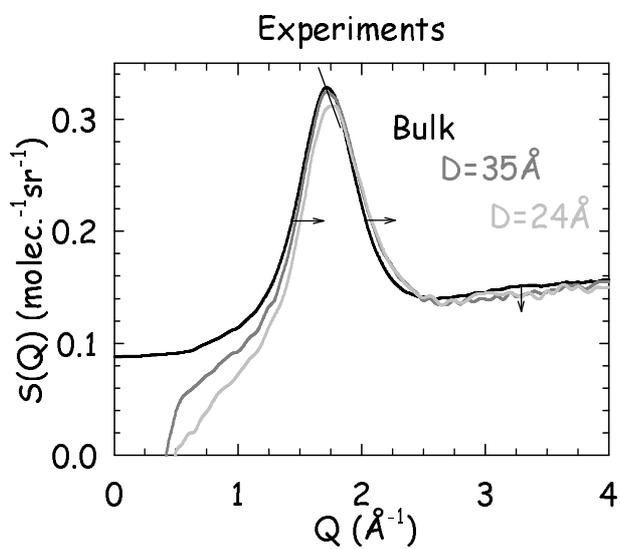

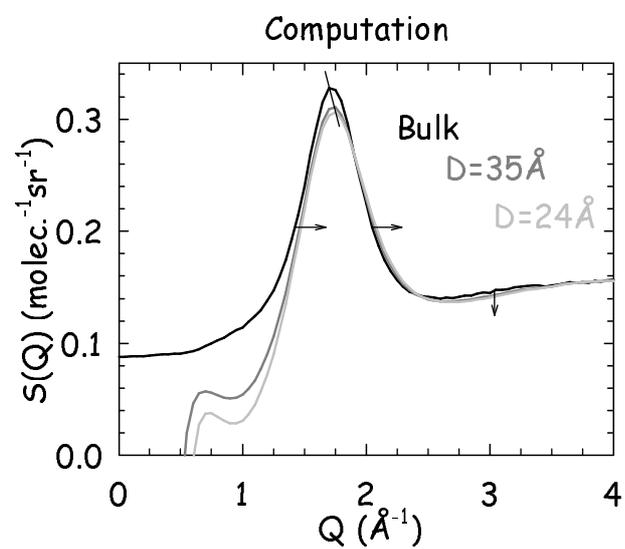

(a)

(b)

Figure 10
**Structure of liquid and glassy methanol confined in cylindrical pores.**
D. Morineau et al.

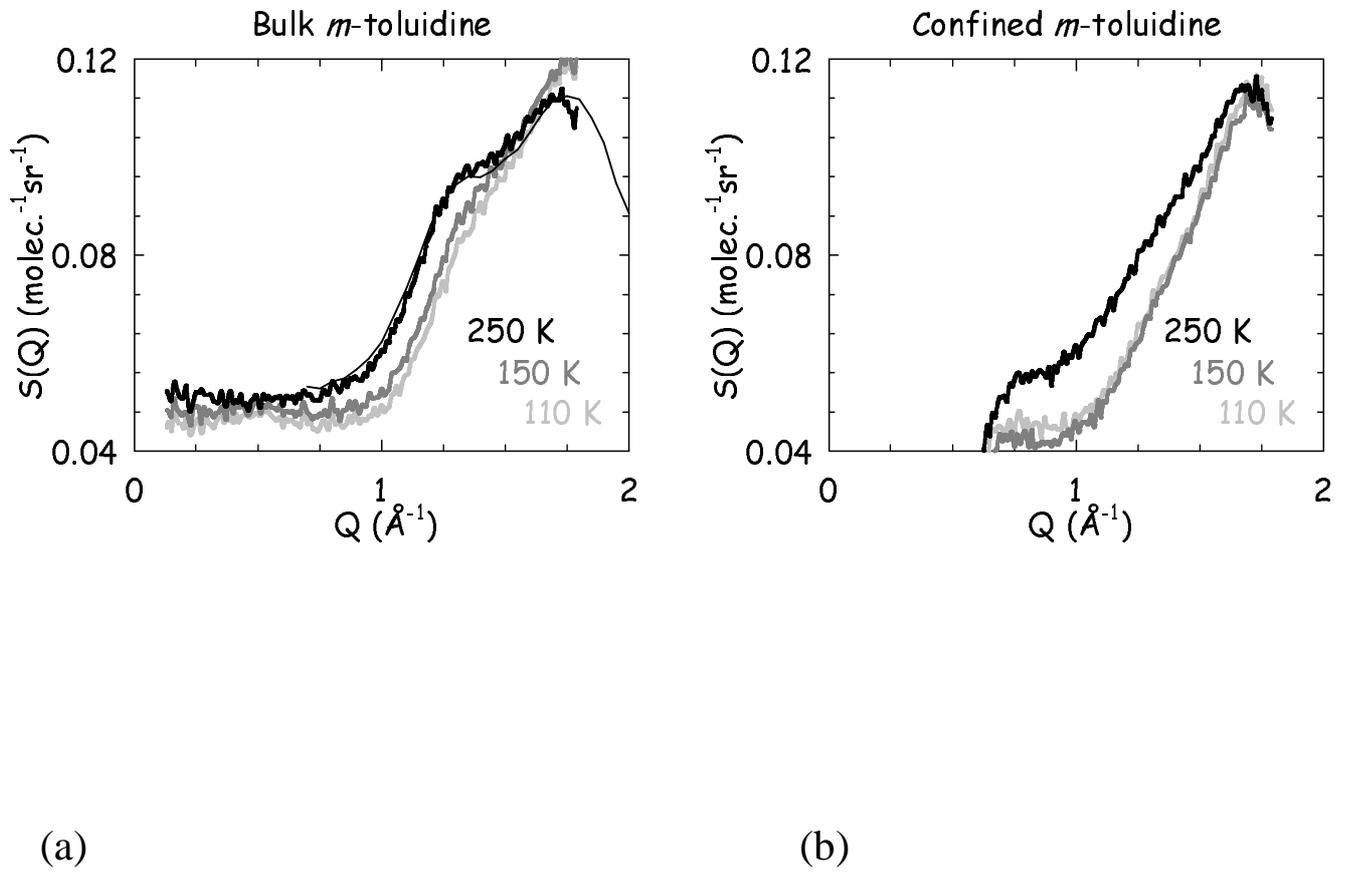

(a)          (b)

Figure 11
**Structure of liquid and glassy methanol confined in cylindrical pores.**
D. Morineau et al.